# Anomalous Effects in Air While Cooling Water

Rachel Sardo* and James D. Brownridge**
State University of New York at Binghamton
Department of Physics, Applied Physics, and Astronomy
Binghamton, NY 13902

Water is a unique compound with many anomalies and properties not fully understood. Designing an experiment in the laboratory to study such anomalies, we set up a series of experiments where a tube was placed inside a sealed container with thermocouples attached to the outer surface of the tube and in the air adjacent to the tube. Alternately, deionized water and other compounds were added to the tube and cooled to freezing. Several of the thermocouples suspended in the air and adjacent to the tube showed thermal oscillations as the overall temperature of the container was decreasing. The temperature of the thermocouples increased and decreased in a sinusoidal way during part of the cool down to freezing. Thermal oscillations as large as 3ºC were recorded with typical frequencies of about 5 oscillations per minute.

**1.0.0 Introduction**:

Many of water's unique features are still being explored[1,2]. It is the hydrogen bonding of water molecules to other molecules and liquids that make the water molecules unique with properties unlike any other liquid[3]. The temperature differences between the boiling point and freezing point of $H_2O$ is greater than that of other liquids. Based on the molar mass of this molecule, water exhibits a number of unique proprieties. There is an attraction between the water molecules as a result of dipole-dipole interactions. Hydrogen bonding results in a tetrahedral arrangement of hydrogen atoms around the oxygen with two covalent bonds and two hydrogen-bonded hydrogen atoms. The nature of a water molecule and the hydrogen bonding leads water molecules to stick to each other and also to have an effect on the properties of compounds.

As frozen water is heated, the tetrahedral lattice structure breaks down and, contrary to the melting of most other compounds, the mean distance between molecules actually decreases. Temperature affects the hydrogen bonded molecules and the water becomes denser as it is heated up from 0ºC to 4ºC[4]. Because of the nature of water, there are many anomalies. Water clusters, also referred to as 'dynamic heterogeneities', are considered to be a source of some of these anomalies[5]. The temperature dependence of density of water results in another anomaly: turnover. Turnover can be seen in any body of water. Water is at its maximum density at 4ºC, so when the water cools, the water that is less than 4ºC rise while the denser water sinks. The water and dissolved materials in it turnover until it all reaches 4ºC. The cooler water then remains on top, since it is less dense, as these bodies of water freeze.[6]

**1.1.0 Water Clusters**

Water clusters are thought to be present in liquid water. The existence of water clusters theoretically can explain much of the anomalies of water, such as: density, compressibility, heat capacity, and the dielectric constant.[7,8] Through the breaking and bonding of hydrogen, water clusters are formed and reformed.[9] Hydrogen bonds vary in strength; this allows the hydrogen of one water molecule to bond with the oxygen of another water molecule.[10] There are many different theories on what the structure of these water clusters could be.[11]

**1.1.1 Water Cluster History**

In 1884 the first mention of clusters was made by H. Whiting in his Harvard

Physics Phd. Thesis.[12] Whiting studied melting ice. He wrote that low density solid particles that were released during melting had an affect on temperature and pressure. He was the first to write about these "solid particles."

In 1933 Bernal and Fowler were the next to write significantly about the arrangement of water molecules in the liquid state. While looking to explain anomalous changes in the volume of water, they described water molecules in the liquid state to be "quartz-like with appreciable tridymite-ice-like tendencies.[13]" They wrote "there are three chief forms of arrangement of the molecules of $H_2O$ in water: *water I*, tridymite-ice-like, rather rare, present to a certain degree at low temperatures below 4ºC: *water II*, quartz-like, predominating at ordinary temperatures: *water III*, close packed, ideal liquid, ammonia-like, predominating at high temperatures for some distance below the critical point at 374ºC.[14]" The quartz-like structure resembles an ice-like structure of water.

Bernal and Fowler[13] proposed that at different temperatures the structures change. For example the quartz-like structure of water II can become more tightly packed moving into the water III structure. Water II was the most common arrangement in normal water. The only problem with their findings was that they only applied to homogenous liquids.

In 1946 the idea of interstitial water molecules inside an ice hexagonal box was proposed by O. Ya. Samoilov. He thought that there were water molecules inside the ice hexagon, essentially filling up the hole.[15] Interstitial water molecules are basically hydrogen-bonded water molecules with holes in them where another water molecule could fit[16]. In 1959 Pauling proposed another interstitial water molecule model which was based upon the crystal structure of chlorine. It was known as the clathrate model.[17] These two examples of cluster arrangements are not used today. Both arrangements are still to be proven correct, but further development of these ideas is possible.

Bernal later in 1975 discovered another model that could be applicable.[18] Bernal modeled the results of x-ray diffraction and neutron scattering on amorphous solid water. The results from the modeling were compared with experimental results. Bernal proposed that water had different-sized bonded water molecules making a ring. The rings were of many different sizes (consisting of 4 to 8 water molecules bonded together). This model can only be used for homogenous cases.

Robinson et al. published papers explaining anomalies of water and established the "outer structure two-state mixture model."[19] Robinson et al. believed that the structure of liquid water must be related to ice 1h (hexagonal ice structure of water molecules), ice II (rhombohedral structure of water molecules), and ice III (five-membered ringed structure of water molecules) to explain such anomalies such as the density maxima of $H_2O$, the effect of pressure on the density maxima, and the isotope effects on viscosity.[20] They thought that there was outer oxygen to oxygen bonding. Supposedly, there was always a tetrahedral structure. Still, it was not known whether water clusters with ices 1h, II, and III could be used for liquid water.

Doherty and Howard proposed the equilibrium model, in 1998.[21] This model explained anomalies of water such as density, heat capacity at constant pressure, heat capacity at constant volume, isothermal compressibility, viscosity as a function of pressure and temperature, acidity, heat of fusion, and hydrocarbon solubility. As the temperature increased, there were equilibrium shifts and water clusters became less structured. "The equilibrium structural model for liquid water describes the liquid as a random structural network that is held together by hydrogen bonds.[22]" This model suggested that water clusters consist of 5- or 6- members, making them dodecahedra. This model helped Martin Chaplin to develop his model of clusters structured as icosahedra[23].

Over the past decade, Martin Chaplin has attempted to explain the anomalies of water. In 2000 he published a paper called "A proposal for the structuring of water". Chaplin states, "[water] remains an anomalous liquid where no single model is able to explain all of its properties.[24]" Chaplin used past models to work with and build from to form his model for the structure of water. "The basis of this model is a network that can convert between lower and high density forms without breaking hydrogen bonds.[25]" The model was folded over and essentially was a form of an icosahedral three-dimensional network. It was important for Chaplin to have a model that was semi-collapsible because of the competition between non-bonded and bonded molecules.

**1.1.2 Equipment and Methods of Studying Water Clusters**

There are many methods that can be used to study the structure of water such as: dielectric spectroscopy, x-ray and neutron diffraction, computer models, and nuclear magnetic resonance. Some of these methods require model-based interpretation. The interpretation utilizing theoretical analysis leads to subjective and model-based results.

Dielectric spectroscopy allows one to see the amount of different clusters. There are different types of clusters due to the different hydrogen bond strengths. Each molecular interaction is associated with specific electrical charge distribution. "Dielectric spectroscopy proceeds from the dominant role which electrical charges play in the molecular interactions of condensed matter."[26] Dielectric spectroscopy also allows one to measure water tumbling. When studying a spectrum, the data is usually graphed in terms of frequency versus dielectric loss. If there is a cluster that is tightly bonded, then the water tumbling is slow; whereas, if a cluster is loosely bonded then the water tumbling is fast.

Diffraction methods do not show the geometry of clusters, but these diffraction methods do show sensitivity to electron distribution and show areas with higher electron densities. Diffraction methods, such as X-ray diffraction, do not directly establish a molecular model. These methods can test a molecular model already proposed. The time-averaged distances between water molecules can be found.

Models of water's molecular structure are produced on a computer. Computer models create different simulations and show whether or not the model works in those simulations. Anything can be tested, but the computer simulations rely as much on the scientist choosing initial conditions as to what types of theories and equations will be used in the simulation.

Nuclear magnetic resonance can be useful to study water clusters since water has three isotopes, two isotopes of hydrogen ($^1$H and $^2$H) and one isotope of oxygen ($^{17}$O).[27] The problem with using NMR is that the magnetic fields utilized in NMR can change the water clusters.

Vibrational spectra are used to study water clusters. Water is a strong infrared absorber. The greenhouse effect is in part caused by this absorption.[28] Water vibrates in the liquid phase with the combination of three different modes such as symmetric stretch, asymmetric stretch, and bending. Water has infrared and Raman spectra which have been applied to many different fields such as atmospheric science, astronomy and combustion.[29]

X-ray spectroscopy uses high energy photons to excite electrons. These excited electrons move to unoccupied valence levels. This produces signals that show information about the hydrogen bonding of water.

**1.1.3 Cluster Size**

Giovambuttista, Buldyrev, Starr, and Stanely studied how the cluster size related to temperature and time. They found that as temperature decreased the clusters increased.[30]

Canpolat, Starr, Scala, Lahijany, Mishima, Havlin, and Stanely studied the relation of heterogeneities and pressure. "…when pressure increases; we expect the number of non-hydrogen-bonded pentamers (high density structural heterogeneities) to increase and the number of bonded pentamers (low density structural heterogeneities) to decrease.[31]" If water was subjected to high pressure, there would be a high density of heterogeneities in the water.

**1.2.0 Heavy water**

Water that contains the isotope deuterium is known as heavy water or deuterium oxide. Hydrogen atoms all contain one proton. The isotope of hydrogen called deuterium contains one neutron. Although heavy water looks and tastes like light water or $H_2O$, the properties are different. The freezing point is higher at 3.82ºC, yet the boiling point is 101.4ºC, which is close to $H_2O$'s boiling point of 100ºC. It is the density of the heavy water that makes the properties different from $H_2O$. A heavy water ice cube will sink in $H_2O$. Heavy water is 10.6% more dense than light water or $H_2O$. In large doses heavy water is harmful to humans.

**2.0.0 Experimental Procedure**

The experimental set up was developed as a result of the need to create a system that was simple and unproblematic when changing the parameters. The experimental set up was used to measure the temperature of the air in a container just outside a tube and it was also used to measure the temperature of the water inside the tube based on the experiment that was being done. The experimental set up involved a container, a tube hanging from the lid of the container, and thermocouples set up at different distances in the air inside the container. The container was cooled by placing it inside a freezer. Figure 2.1 illustrates the container with thermocouples coming out of the container connected to the Omega Data Acquisition Hardware, which was then connected to the computer. It was necessary to have a way of reading the temperature of the air inside the container. Thermocouples were used to record the temperature of the air. Omega Data Acquisition software and hardware was used to interpret the thermocouple readings. Throughout all experiments, there was a thermos filled with an ice water bath with a thermocouple inside which was connected to the Omega Data Acquisition Hardware. The ice water bath served as the reference temperature throughout these experiments and showed that during most experiments the temperatures read by the other thermocouples were offset by around 2.5ºC.

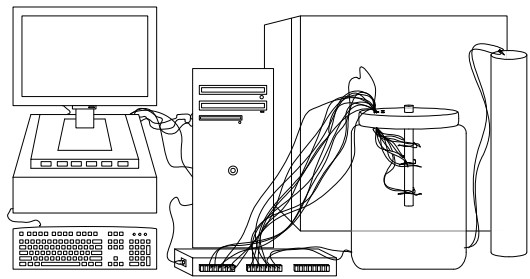

**Figure 2.1:** This was the experiment set up when experiments were done. Omega Data Acquisition Hardware was used. A container had thermocouples that were placed at different distances away from the tube inside the container. Also, one thermocouple was always placed in a thermos which was to the right of the set up. There was a freezer used to cool the container.

**2.1.0 The Container**

The container used in the experiment was a cylindrical container with a height of 24cm and a diameter of 10cm (Figure 2.2). A hole was drilled into the center of the lid; and the lid was then sealed onto the container. This hole had the same diameter as the tube that was to be used throughout the experiment. The hole was drilled in such a way that it was loose enough so that the tube could fit and hang from the lid, but tight enough so that it would not fall out. Two other holes about 1mm in diameter were drilled into the lid of the container for the thermocouples.

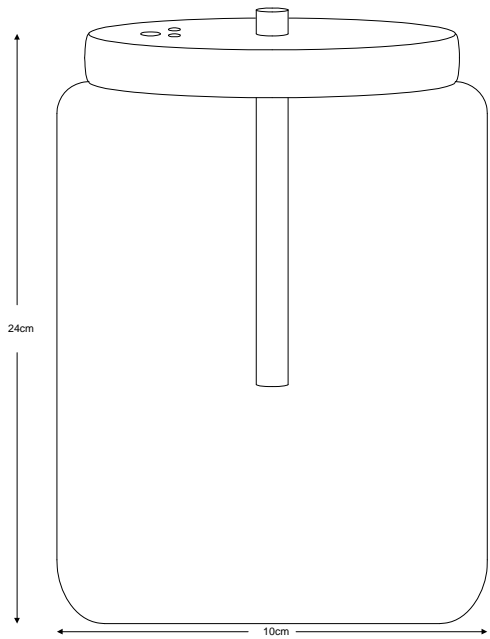

**Figure 2.2:** The dimensions of the container used in all experiments.

The tube hanging from the lid had either different liquids in it or no liquid at all. In order to keep the tube insulated, the tube was always plugged with a foam-like material that was rolled up to fit inside the diameter of the tube. Thermocouples were placed in the air at varying distances away from the tube and also on the tube as shown in Figure 2.3.

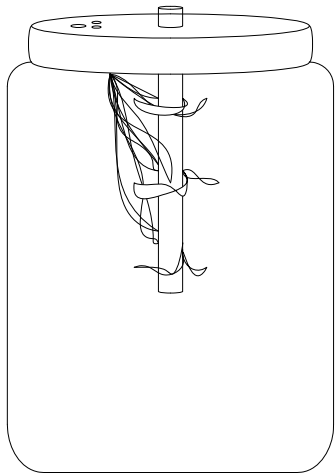

**Figure 2.3:** This was just one example of the container. In this diagram there were 6 thermocouples hanging around the tube.

At the start of each experiment, the freezer was cooled to a temperature of approximately -17ºC. The tube was then filled at various times with water, alcohol, or left empty. The thermocouples were placed in positions as specified above. The container was then sealed and insulated. The container was subsequently placed into the freezer and cooled below 4ºC. The amount of time that the container was left in the freezer changed from experiment to experiment depending on the size of the tube and the type of experiment that was being performed. For example, when the experiment was designed to demonstrate latent heat release and freezing, the amount of time the container needed to be in the freezer was lengthened. Once in the freezer, the Omega Data Acquisition Program was used to measure the temperature of the air outside the tube as time progressed. Oscillations in temperature were sometimes seen. Described as sinusoidal variations, these oscillations would occur as the temperature decreased.

### 2.2.0 The Tubes

Only three different size tubes were used in these experiments. It was important to try to keep the material of the tubes the same. Two more plastic tubes of the same material were used, one tube with a 7mm diameter and the other with a 2.1cm diameter. A cylindrical piece of wood measuring either 7mm in diameter or 2.1cm in diameter was used to make two of the tubes. A piece of Teflon was wrapped around the wooden cylinder. Thin wall polyolefin heat shrink tubing encased the Teflon and wooden cylinder. The shrink tubing was then heated and shrunk to the diameter of the wood. The Teflon prevented the shrink tubing from sticking to the wood and facilitated the removal of the shrink tubing from the wood after the Teflon was removed. The 7mm diameter tube and the 2.1cm diameter tube were cut to a length of approximately 13cm, which was the same as that of the 5cm diameter tube. In order to ensure that one side of the tube was permanently closed, about 1.5cm of epoxy was applied. For comparison, one plastic tube with a 5cm diameter and a length of 13cm was used, but was not made with shrink tubing. Instead, it was a commercially plastic tube with a sealed bottom.

### 2.3.0 Thermocouples

The thermocouples were used to measure the air temperature inside the container at different heights, as well as to read the temperature just outside the tube at different heights. The temperature of the thermocouples taped to the tube was close to the temperature of the water inside the tube. The thermocouples that were used for these experiments had to be hand welded. They were T-type thermocouples with a diameter of .07mm (.003"), which made the welding difficult.

The T-type thermocouple was made from Copper and Constantin wire. The insulation of the wires was color-coded making the Copper blue and the Constantin red. In order to weld the thermocouples together, the Copper and the Constantin wires were twisted together. Using a small welder, the twisted thermocouple was then placed as close to the hottest part of the flame as possible and removed as soon as there was a bead produced by the Copper. The remaining part of the wire was untwisted so that all that was left was the bead joining the Constantin to the Copper. Had the wires been touching in other places, there would have been more than one signal coming into the thermocouple. We did this to eliminate systematic errors from the thermocouple measurements.

The thermocouples needed to be placed right outside the hottest point of the flame, otherwise, the Copper would oxidize and the thermocouples could not be welded properly. Another reason for caution was that if the thermocouples were left in the flame for too long, the Copper would melt away and the Constantin would break off, thus ruining the connection.

### 2.4.0 DAQ

The thermocouple wires left the container through the small holes that had been drilled through the lid. These wires were connected to the Omega Data Acquisition (DAQ) Hardware. The DAQ Hardware that was used for these experiments was the Personal Omega DAQ-3000 model. The Omega DAQ-3000 was connected to the computer. This way the information goes straight from the DAQ-3000 to the computer. The software, DaqView, was loaded on to the computer in order to help analyze and view this data. There were a total of 8 channels on the DAQ Hardware that could be used. For one run of the experiment there can be at most 8 thermocouples used. One of these channels was always used to reference all the data to the ice water bath outside the freezer.

### 2.5.0 Grounding the Omega DAQ 3000

To prevent random spikes in the temperature on any run from occurring, the DAQ was grounded. The spikes were eliminated so that all data collected was dependable. The ice water bath was steady at -2.5 degrees Celsius. The temperature was offset by 2.5 degrees.

### 2.6.0 The Freezer

The Freezer that was used in these experiments was a Holiday Refrigerator adjusted to function as a freezer. Since the freezer needed to have a uniform temperature throughout, a 6V fan was placed in the freezer compartment of the refrigerator in order to distribute the temperature evenly throughout the freezer. Convection currents produced by the fan produced a uniform temperature throughout the freezer.

### 2.7.0 Insulation

It was determined that the container needed to be better insulated to slow the rate of cooling and to enhance oscillations. Prior to the insulation, the oscillations, when they did occur, were very weak. Insulating the container helped the container cool at a slower rate. We found that the rate of cooling directly affected the oscillations. The container was insulated by wrapping a foam-like plastic material around the whole container so that it was fully insulated.

Since adding one layer of insulation worked so well in the first few attempts, the cooling rate was slowed down even more by wrapping yet another extra layer of insulation around the container. There was also another layer that was wrapped around the top of the insulation that was again secured with a rubber band, thus creating an even more uniform temperature. As results later showed, two layers of insulation did not produce large oscillations

**2.8.0 Water in Tubes**

All water used was deionized water. The water was carefully added to various size tubes by using a syringe. A tube with a diameter of 3mm and a length of 15cm was attached to the syringe (Figure 2.4). The syringe with the small diameter tube was used to suck up the deionized water. The deionized water was then pushed into the tube, using the syringe of deionized water. This was done carefully so as to not create any air bubbles.

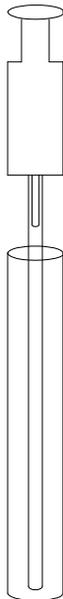

**Figure 2.4:** The syringe shown above needed to be used to insert deionized water into the tube.

In order to have as few variables as possible, air bubbles were kept to a minimum. Air bubbles may directly prevent or cause oscillations. The air bubbles in some way contributed to the experimental results.

**3.0.0 Results/Discussion**:

In all cases, T-type thermocouples with a width of 0.07mm (0.003'') diameter were used. Tubes of different diameters all had a length of 13cm. These tubes were hung from the lid of a container with a height of 24cm and a diameter of 10cm. One centimeter of tube protruded from the lid. The container was surrounded by insulation. Before each experiment the freezer was on until it reached its coldest temperature, about -17ºC. The container was always insulated. All experiments were run with one thermocouple in an external ice water bath to measure any electronic drift. The ice water bath was always about 2.5ºC lower.

**3.1.0 Room Air vs. Water**

The question arose, what would happen if the tube was empty? The tube was empty for the series of experiments labeled 3.1.1 to 3.1.4. Water was added to the tube for the experiments from 3.1.5 to 3.1.7. These experiments showed that the oscillations that occurred in the air had a greater temperature swing when there was water inside the tube. However, oscillations in the air could still occur with an empty tube.

**3.1.1 Case19feb07c: Air in a Tube with a Diameter of 7mm**

A tube with a diameter of 7mm was used in this case. There was nothing placed in the tube. The tube was sealed. The only thing inside the tube was room temperature (25ºC) air. The thermocouples were arranged to be different distances from the tube, but they were all located about halfway up the length of the tube. Figure 3.1 showed the distances of the thermocouples away from the tube. The colors correspond to the graphs of Figures 3.2 through 3.22. The thermocouple touching the tube was also halfway up the length the tube. The other thermocouples were 3mm away, 5mm away and 9mm away, all halfway up the tube. The thermocouple on the outside of the container was about halfway up the container, which would make it about 12cm above the table if the container was resting on the table. As always there was an ice water bath for thermocouple reference.

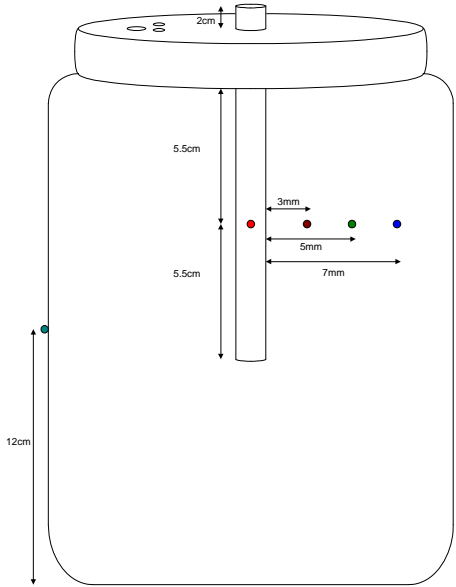

**Figure 3.1:** The colored dots inside and outside the container correspond to the curves representing thermocouples in graphs for Case19feb07c, Case19feb07e, Case19feb07g, Case20feb07a, Case20feb07c, Case20feb07e, and Case20feb07g.

After the thermocouples were arranged, the lid was sealed, and the container was insulated. The container was placed in the already cold freezer and cooled. The container was in the freezer for about 43 minutes. There were 10,929 data points taken at a rate of about 5 pts. per second.

Figure 3.2 showed the temperature decrease as it was cooled. When examined closer from a time scale of 6-12 minutes, one can see the oscillations in the graph (Figure 3.3). These oscillations lasted about 4 minutes and disappeared. Fluctuations were then seen at around 11 minute into the run. These fluctuations did not have a clear pattern and did not last long.

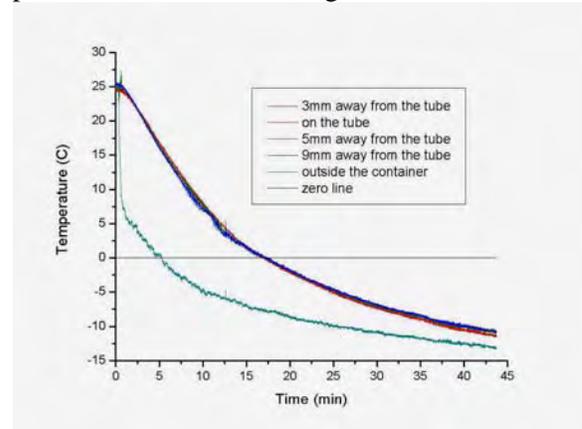

**Figure 3.2:** Temperature vs. Time graph of the container cooling down (Case19feb07c). Only air was present inside the tube.

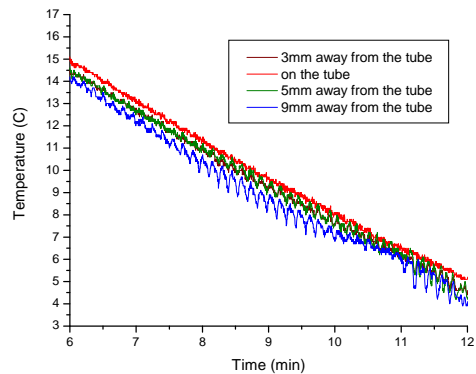

**Figure 3.3:** Temperature vs. Time graph for Case19feb07c from 6 minutes into the run to 12 minutes in.

In this case, the thermocouple touching the tube (red) showed no sign of oscillations. The air around the thermocouples that were 3mm away (brown) and 5mm away (green) were oscillating in phase with each other and with the same amplitude. The air temperature around these two thermocouples fluctuated by 0.5 of a degree. The temperature of the air oscillated with a period of 9.12 seconds. The air around the thermocouple that was 9mm (blue) away also had a temperature that varied in phase with the air around the brown and green thermocouples. Figure 3.4 showed the run from 480 seconds to 570 seconds with two vertical black lines to show how the air temperature change was in phase. Although the thermal oscillations in air around these three thermocouples had the same period, the air around the blue thermocouple had greater amplitude by 0.5 of a degree. The thermal energy associated with the "thermal bubble" was greatest in the area of the blue thermocouple since the temperature change was greatest.

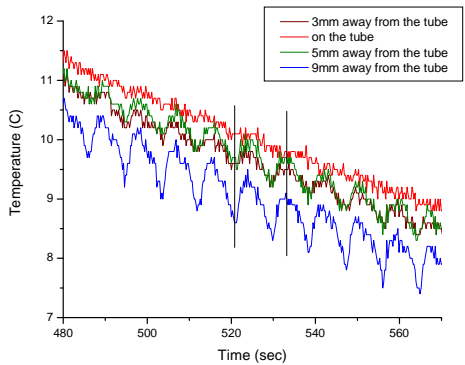

**Figure 3.4:** Temperature vs. Time graph for Case 19feb07c over a time interval from 480 seconds to 570 seconds. The black lines on the graph show how the thermal oscillations in the air just outside the tube at different positions were in phase with each other.

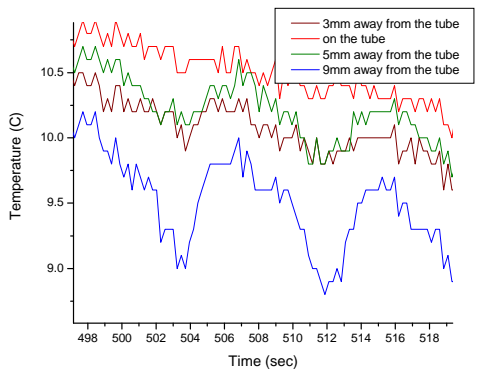

**Figure 3.5:** Temperature vs. Time graph for Case19feb07c over the time interval from 497 seconds to 519 seconds.

### 3.1.2 Case19feb07e: **Air in a Tube with a Diameter of 7mm**

This case was the same as Case19feb07c. Nothing was changed. All the thermocouples remained in the same positions which can be seen in Figure 3.1. The only time the container was touched was when the container was removed from the freezer at the end of Case19feb07c. The container and air within it remained outside the freezer until both came to room temperature. The container was then put back into the freezer and cooled for 24 minutes. Figure 3.6 showed the temperature decreasing with time. A closer look over the time period of 12 to 15 minutes showed that there were no oscillations during this run (Figure 3.7). There were fluctuations, but there was no pattern. All parameters that could be controlled remained the same, but the results differed completely. Case19feb07c showed oscillations and Case19feb07e just showed fluctuations.

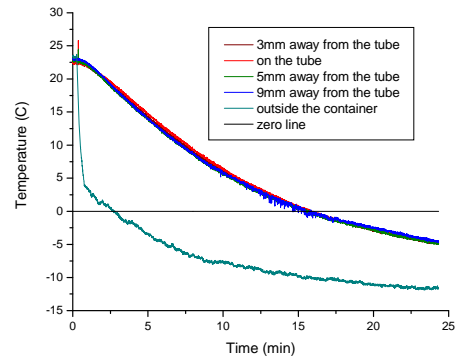

**Figure 3.6:** Temperature vs. Time graph of the container cooling down (Case19feb07e). Only air was present inside the tube.

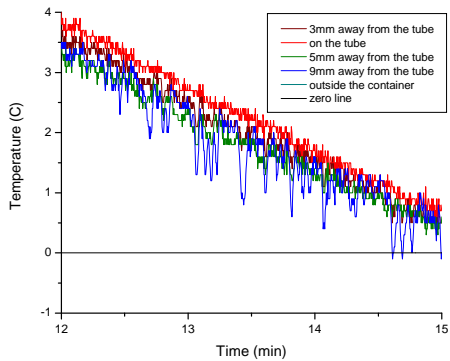

**Figure 3.7:** Temperature vs. Time graph of Case19feb07c over the time interval from 12 minutes into the run to 15 minutes in. Thermal fluctuations were seen, but no thermal oscillations.

### 3.1.3 Case19feb07g: Air in a Tube with a Diameter of 7mm

This case was the same as Case19feb07c and Case19feb07e. Again, nothing was changed. The thermocouples were in the same position as Figure 3.1. Nothing was touched, except for taking the container in and out of the freezer. The container was in the freezer for around 28 minutes. Figure 3.8 showed the decrease in temperature with some oscillations. Figure 3.9 showed that the oscillations began at around 10 minutes into the run and lasted for about 5 minutes. The thermocouple touching the tube (red) did not oscillate at all.

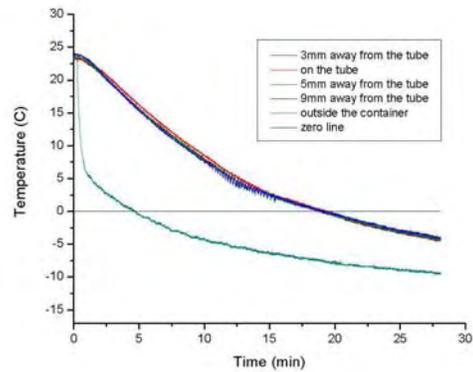

**Figure 3.8:** Temperature vs. Time graph of the container cooling down (Case19feb07g). Only air was present inside the tube.

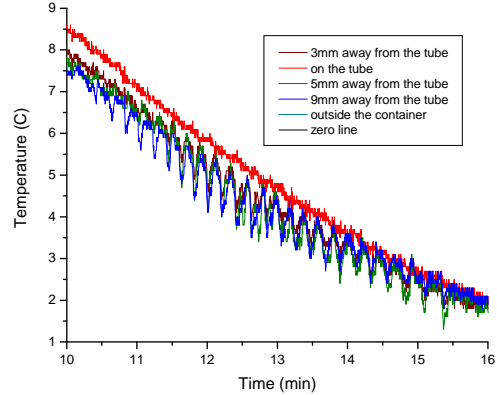

**Figure 3.9:** Temperature vs. Time graph for Case19feb07g over the time interval from 10 minutes into the run to 16 minutes in. Thermal oscillations were seen.

The data is presented so that the readings of each thermocouple in air could be seen better (Figure 3.10). The thermocouple 3mm away was not changed, but temperature of the thermocouple 5mm had one degree taken away so it moved down on the graph and the thermocouple 9mm away had two degrees taken away. The thermocouples registering the air temperature all oscillated at the same rate and fluctuated about 1 degree.

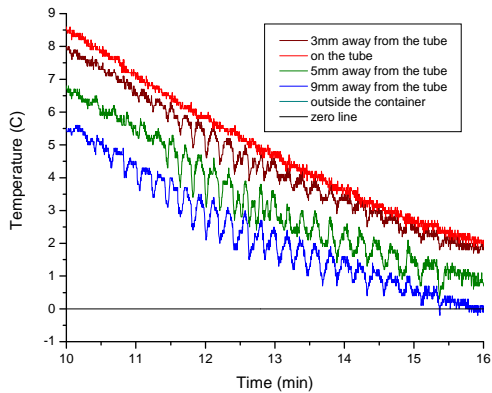

**Figure 3.10:** Temperature vs. Time graph of Case19feb07g. Figure 3.9 was manipulated to separate the curves of the three thermocouples oscillating.

### 3.1.4 Case20feb07a: Air in a Tube with a Diameter of 7mm

This case was the same as Case19feb07c, Case19feb07e, and Case19feb07g. The container was only handled when taken out of the freezer after Case19feb07g was finished. Once the temperature of the air inside the container reached room temperature, it was put into the freezer. There it cooled for about 35 minutes (Figure 3.11). Figure 3.12 showed the time interval where the temperature oscillations took place (between 7 and 16 minutes).

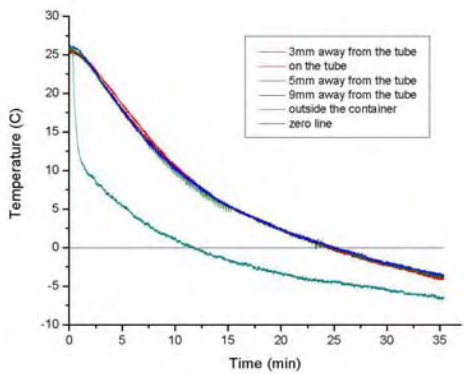

**Figure 3.11:** Temperature vs. Time graph of the container cooling down (Case19feb07a). Only air was present inside the tube.

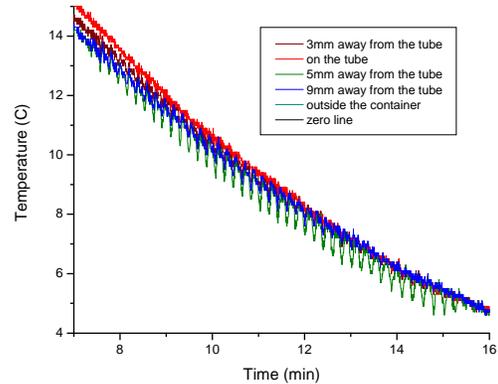

**Figure 3.12:** Temperature vs. Time graph for Case20feb07a over the time interval from 7 minutes to 16 minutes.

When examining these oscillations closer, one can see that the three thermocouples reading the temperature of the air are in phase with each other (Figure 3.13). The thermocouple on the graph that was 3mm away (brown) was moved down a half of a degree by subtraction (Figure 3.14). This was also done with the two other thermocouples that were in the air; the thermocouple 5mm away (green) was moved down 1 degree by subtraction; and the thermocouple 9mm away (blue) was moved down 2 degrees by subtraction. The region with the strongest oscillations, the largest temperature change, was the region 5mm away from the tube. There was a 1 degree swing for the air that was 5mm away, a 0.3 degree swing for the area 3mm away, and a 0.5 degree swing for the area 9mm away. The air temperature was greatest between the outermost thermocouple and the thermocouple closest to the tube, but not in direct contact with it.

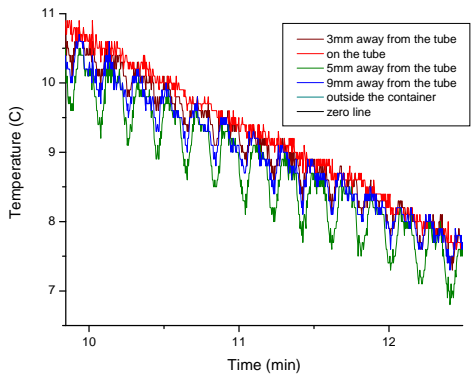

**Figure 3.13:** Temperature vs. Time graph for Case20feb07a over the interval from 10 minutes to 12 ½ minutes.

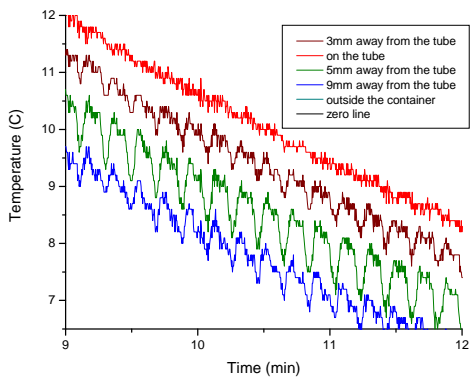

**Figure 3.14:** Temperature vs. Time graph for Case20feb07a over the interval from 9 minutes to 12 minutes. The curves were separated in order to show the three positions inside the container that had thermal oscillations.

### 3.1.5 Case20feb07c: Deionized Water in a Tube with a Diameter of 7mm

This case was the same as Case19feb07c, Case19feb07e, Case19feb07g, and Case20feb07a except that 28mL's of deionized water (with a temperature of 23ºC) were pushed into the tube using a syringe with a smaller tube attached to it. Figure 3.1 showed the way the thermocouples were arranged for these cases. The lid to the container had been sealed since Case20feb07a was finished and did not need to be opened to put the water inside the tube. Once the water was inside the tube, the tube was plugged and the insulation was put around the container again. The container was then cooled for 47 minutes (Figure 3.15). Figure 3.16 showed that the oscillations originated after 5 minutes and lasted for 12 minutes.

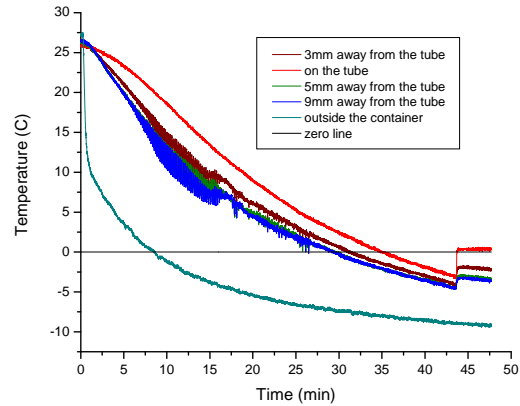

**Figure 3.15:** Temperature vs. Time graph of the container cooling down (Case20feb07c). Deionized water was inside the tube.

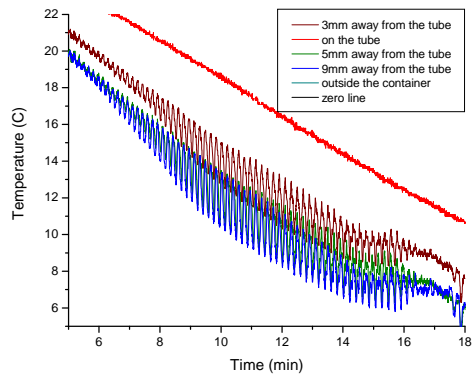

**Figure 3.16:** Temperature vs. Time graph of Case20feb07c over the interval from 7 minutes to 18 minutes.

Figure 3.17 showed clearly that the three thermocouples with oscillations on them were in phase with each other. One degree was added to the thermocouple that was 3mm away (brown) and two degrees was taken away from the thermocouple which was 9mm (blue) away. This adjustment separated the oscillations of the air temperature in different regions so each could individually be seen. For most of the time, the air temperature in all three areas were oscillating with the same period and had the same temperature swing of about 3 degrees.

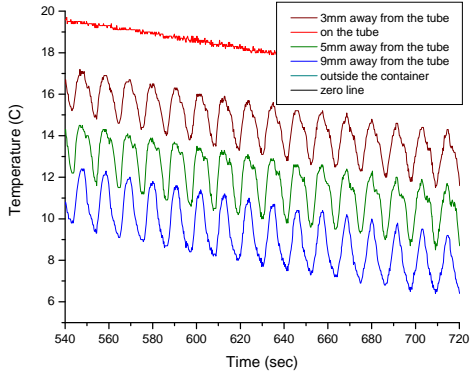

**Figure 3.17:** Temperature vs. Time graph of Case20feb07c over the interval from 540 seconds to 720 seconds. The three curves were separated for a better look at the thermal oscillations.

### 3.1.6 Case20feb07e: Deionized Water in a Tube with a Diameter of 7mm

This case was the same as Case20feb07c. The same water that was used in Case20feb07c was used in Case20feb07e. Nothing was changed. The thermocouples were still arranged in the same way as Case19feb07c through Case20feb07e (Figure 3.1). The container was taken out of the freezer after Case20feb07c finished. The container was brought to room temperature and then placed back in the freezer for Case20feb07e for an hour and 4 minutes (Figure 3.18). Oscillations started at around 5 minutes and lasted around 9 minutes (Figure 3.19). Figure 3.19 also showed that the thermocouples that were 5mm away and 9mm away were both showing the same temperature change; whereas the thermocouple 3mm was slightly higher in temperature.

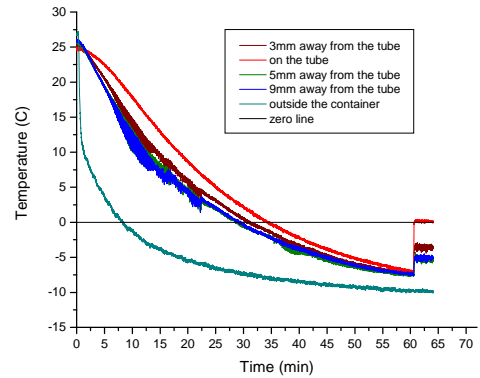

**Figure 3.18:** Temperature vs. Time graph of the container cooling down (Case20feb07e). Deionized water was inside the tube.

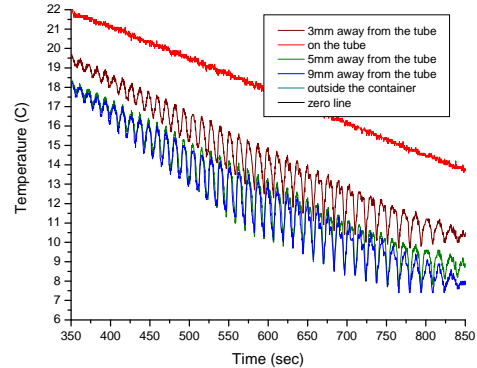

**Figure 3.19:** Temperature vs. Time graph of Case20feb07e over the interval from 350 seconds to 850 seconds.

Figure 3.20 showed that the air temperature around all of the thermocouples was oscillating at the same rate. The thermocouple 5mm away (green) was separated from the 9mm away (blue) thermocouple by subtracting 2 degrees. All three have about the same temperature swing of 3 degrees.

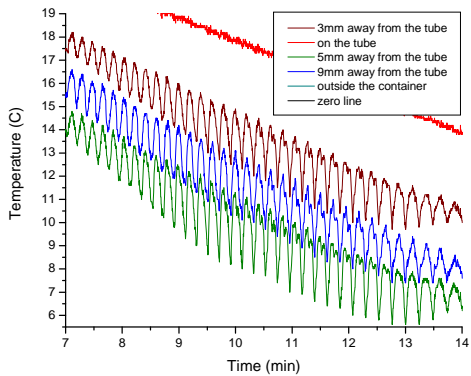

**Figure 3.20:** Temperature vs. Time graph of Case20feb07e over the time interval from 7 minutes to 14 minutes. The three curves were separated for a better look at the thermal oscillations.

### 3.1.7 Case20feb07g: Deionized Water in a Tube with a Diameter of 7mm

    This case was the same as Case20feb07c and Case20feb07e. Again, the same water was in the tube as during the previous two runs. Nothing was altered. After Case20feb07e was finished, the container was taken out of the freezer and, eventually, the air inside warmed to room temperature and that was where Case20feb07c started.

    The container was placed in the freezer and cooled for 58 minutes (Figure 3.21). The oscillations were small and lasted about 7 minutes (Figure 3.22). The oscillations were again the same for the region of air around the thermocouple that was 5mm away (green) and the thermocouple that was 9mm away (blue). The temperature swing was about a 0.5 degree.

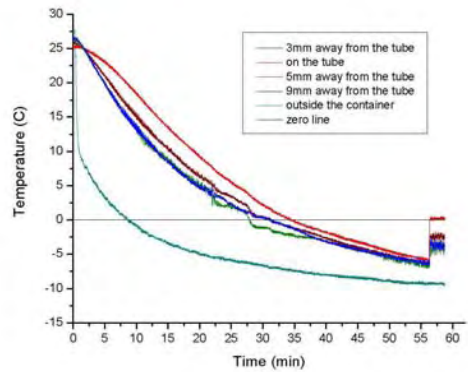

**Figure 3.21:** Temperature vs. Time graph of the container cooling down (Case20feb07g). Deionized water was inside the tube.

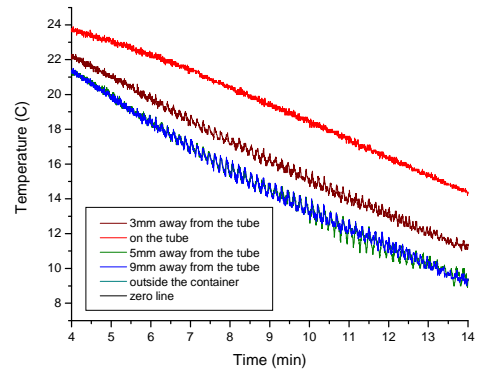

**Figure 3.22:** Temperature vs. Time graph of Case20feb07e over the time interval from 4 minutes to 14 minutes.

### 3.1.8: Summary of Room vs. Air

    During these runs it was important to keep everything constant, except when adding water. It was clearly shown that the oscillations were much larger when water was inside the tube versus just air. The oscillations were greater by about 2.5 degrees with water inside the tube. These runs also showed that everything could remain the same, but an uncontrollable variable might have made a difference and caused the oscillations to become smaller.

### 3.2.0 Different Gases inside the Container

    This series of experiments focused on testing whether or not the type of gas

molecules inside the container contributed to the oscillations that had occurred. The container was adjusted so different gases could be blown in. A hole was drilled into the lid so a straw could be sticking out of the lid. This way a tube could be placed over this straw. The tube led to a balloon which was filled with some type of gas or gas from the nozzle in the lab (which was regular air).

As gas was pushed into the container, there was no tape over the two holes that had thermocouples running through them. The air from the room that was currently inside the container could then leave as the new gas was being pushed in. A balloon was used to transfer the Argon, Helium, and Carbon Dioxide gases. The balloon contained 3.5-4 times the gas volume of the air inside the container; the container held about 754cm$^3$ of the type of gas. In every case the whole balloon of gas was pushed into the container.

Along with testing Argon, Helium, and Carbon Dioxide gases, dry air was also tested. A tube went from the air valve inside the lab room to a cylindrical container that contained Drierite. Another tube came out of the other end of this container with Drierite. The tube covered the straw sticking out of the container that was to be put into the freezer. The air needed to be passed through the Drierite in order to remove all water vapor from the air, thus making the air dry air. The air was turned on for 4 minutes to ensure that all the air in the container was dry air.

### 3.2.1 Case14mar07a: Argon Gas inside the Container

In this case, 14mL of deionized water was inside the 7mm diameter tube. Figure 3.23 showed the arrangement of the thermocouples. The container was filled with Argon gas and cooled for 30 minutes (Figure 3.24). The thermocouple (purple) touching the tube had lost contact with the tube because there was no indication that the latent heat was released at around 25 minutes. The temperature of the thermocouple did not go to zero degrees as it should. Oscillations in the Argon gas temperature occurred after 4 minutes and ranged from 16ºC to 18ºC depending on which thermocouple was studied at that time. These oscillations in the Argon gas temperature lasted for 3 minutes and occurred at three different positions away from the tube.

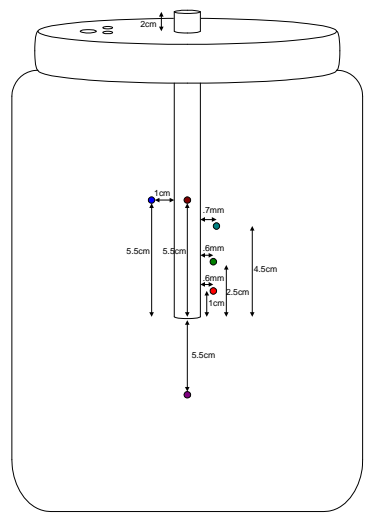

**Figure 3.23:** The colored dots inside and outside the container correspond to the curves representing thermocouples in graphs for 14mar07a, 14mar07b, 14mar07c, and 14mar07d.

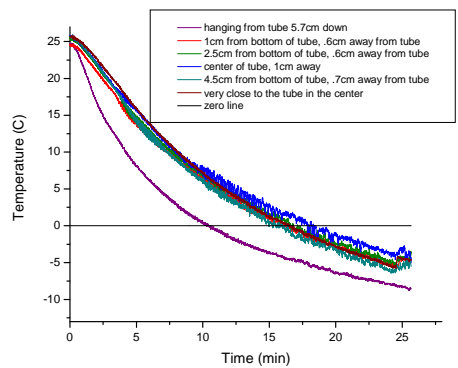

**Figure 3.24:** Temperature vs. Time graph of the container cooling down (Case14mar07a). Argon gas filled the inside of the container. Deionized water filled the tube.

Figure 3.25 was adjusted to these four thermocouples so they could be read separately, and not, overlapping. They were separated by adding or subtracting a few degrees. This case was interesting since oscillations occurred on thermocouples that coincided with Teal, Green, and Red on the graph, but the position that coincided with the blue thermocouple had insignificant activity. After the three minutes of oscillations in the Argon gas, there was 3 minutes of no oscillations in any of the positions. At 10 minutes the position that coincided with the blue thermocouple on the graph began to oscillate for 5 minutes. These oscillations were a little jagged and had a 2ºC temperature swing.

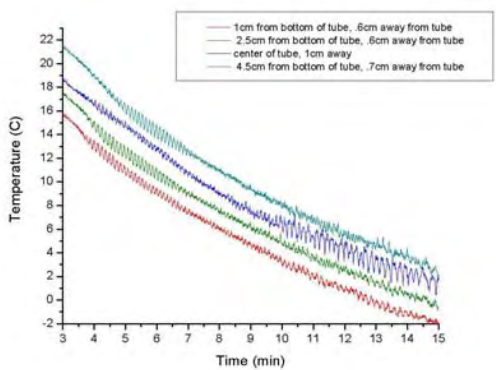

**Figure 3.25:** Temperature vs. Time graph for Case14mar07a over the time interval from 3 minutes to 15 minutes. The three curves were separated for a better view. The "thermal bubble" seemed to have moved from one position to another.

### 3.2.2 Case14mar07b: Room Air inside the Container

In this case air from the room (undried air) was inside the container. There was 14mL of deionized water inside the 7mm diameter tube. The container was in the freezer for 23 minutes (Figure 3.26). Again, the thermocouple we thought in contact with the tube was still slightly displaced from it. The remarkable part of this run was that oscillations occurred in the thermocouple that was 5.7cm below the tube (which is almost to the bottom of the container). This activity had not been seen before. Figure 3.27 showed closer view of these oscillations.

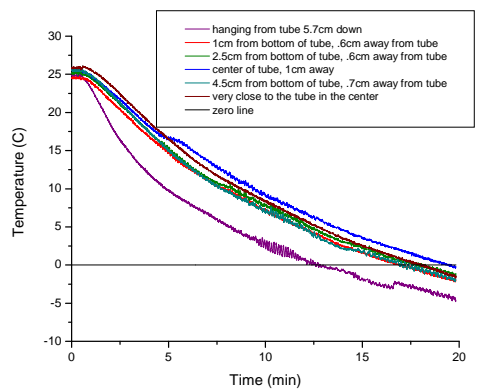

**Figure 3.26:** Temperature vs. Time graph of the container cooling down (Case14mar07b). Room air (undried gas) filled the inside of the container. Deionized water filled the tube.

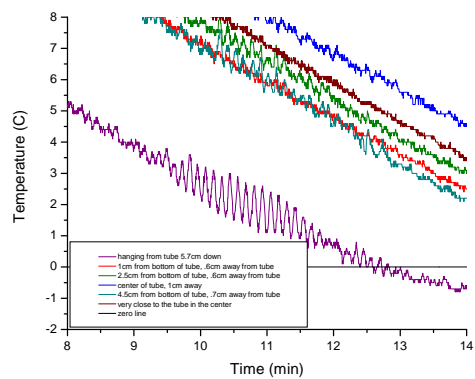

**Figure 3.27:** Temperature vs. Time graph for Case14mar07b over the time interval 8 minutes to 14 minutes. It is clear to that there were thermal oscillations occurring beneath the tube.

### 3.2.3 Case14mar07c: Room Air inside the Container

After Case14mar07b was done, another test using undried room air was completed. The purpose of this run was to see if the oscillations towards the bottom of the container would occur again. Figure 3.28 showed that there were no oscillations in the air on the thermocouple towards the bottom of the container, but there were small ones on two of the thermocouples in the air around the center of the tube and towards the bottom of the tube (Figure 3.29).

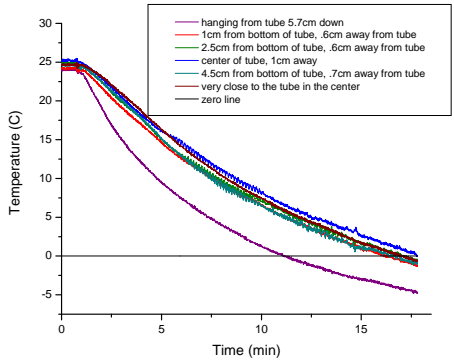

**Figure 3.28:** Temperature vs. Time graph of the container cooling down (Case14mar07c). Room air (undried gas) filled the inside of the container. Deionized water filled the tube.

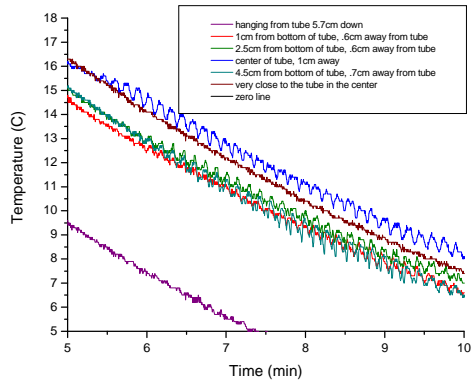

**Figure 3.29:** Temperature vs. Time graph for Case14mar07c over the time interval from 5 minutes to 10 minutes.

### 3.2.4 Case14mar07d: Dry Air inside the Container

The thermocouples for this experiment were arranged the same way as Case14mar07a to Case14mar07c (Figure 3.23). Dry air went into the container for 4 minutes. After the dry air filled the container, the container was placed into the freezer. It was cooled for 20 minutes (Figure 3.30). Fluctuations were seen on all thermocouples except the one closest to the tube halfway up the tube (Figure 3.31).

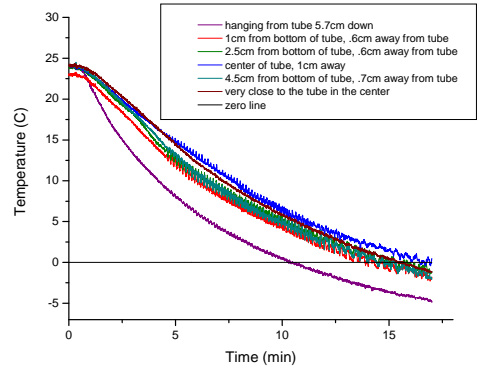

**Figure 3.30:** Temperature vs. Time graph of the container cooling down (Case14mar07d). Dry air filled the inside of the container. Deionized water filled the tube.

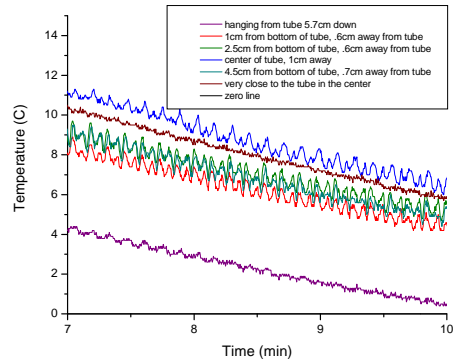

**Figure 3.31:** Temperature vs. Time graph for Case14mar07d over the time interval from 7 minutes to 10 minutes.

### 3.2.5: Summary of Different Gases inside the Container

When Helium and Carbon Dioxide gas were pushed inside the container (in the same way Argon was pushed in) thermal oscillations occurred. The type of gas molecules inside the container did not seem to affect the temperature fluctuations in the gas. The oscillations occurred in all types of surrounding gas. These experiments helped with the understanding of when these oscillations do occur.

**3.2.6: Conclusion**

The anomaly in air occurred when a small diameter tube inside a container was cooled; oscillations in the surrounding gas temperature were seen. The question was posed - what caused this change in gas temperature? We believed that, a temperature gradient inside the container, and the rate of cooling, directly affected these oscillations. We observed that specific conditions needed to be met in order to create these oscillations. There was no way of knowing exactly what the conditions needed to be using the available equipment. All that was known was that the temperature increased by a few degrees and, then, a few seconds later, dropped a few degrees. What was causing the thermocouples to read such a change in temperature?

Oscillations in air temperature that occur inside the container were difficult to explain. Thermals, a column of warm air, may offer an explanation for this anomalous affect. With the equipment available to us for these experiments, there was no way of discovering what really happens inside the container; one could only speculate. A larger sample of about fifty thermocouples would generate a better picture of what exactly was happening inside the container. Constraint on these experiments only allowed seven thermocouples, at most, to be used inside the container. During these experiments we observed "heat" moving from one thermocouple to another.

It is believed that uneven heating in air created thermals. There was a temperature gradient inside the container. The tube (either filled or empty) was the thermal mass inside the container. Convective flows from the thermal mass may create "thermal pockets" or "thermal bubbles" of air in the container. ("Thermal bubbles" are depicted in Figure 4.1). The temperature of the air near the tube was the warmest and the air temperature near the side of the container was coolest. Thermals were formed as a result of this temperature gradient.

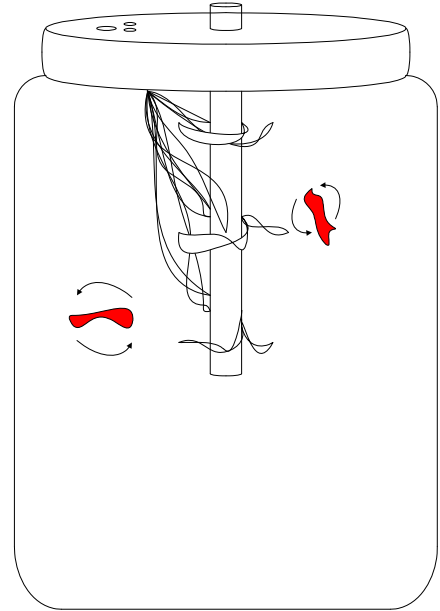

**Figure 4.1:** This figure shows what "thermal bubbles," in red, might look like.

Thermals are used to explain the gliding of engineless planes and birds.[32] When gliders or birds are soaring, thermals need to be found to allow them to maintain altitude. Thermals are formed as the sun heats the ground of a field or some other surface[33]. The ground forms a heat layer just above it and above that there was cool air. The layer of air that was warm rises. As this hump of warm air rose, it sometimes broke away from the ground forming a thermal bubble. Convection caused these "thermal bubbles" to rotate.

Thermals also are very important when studying the weather. It seemed that inside the container, there was a small scale weather system with temperature gradients and thermals. Ziegler[35] and others used computer simulations accordance with observations on the weather of three different days (in May of 1991) to understand drylines and dryline convection. Drylines occurred between dry air and moist air masses.[34] In the Month Weather Reviews of June 1997 they wrote of "thermal bubbles". Thermal bubbles could lead to dryline convection.

Thermals, "thermal bubbles" and drylines could be an explanation for the oscillations in temperature inside the container. If all conditions in this experiment with the equipment used were met, there would be oscillations. It was observed that one condition that must be met was the correct cooling rate. If the container cooled too rapidly or too slowly, oscillations would not occur.

The explanation may be connected to warm pockets of air inside the container. These pockets of warmer air were formed if the container cooled at the correct rate and if there was a large difference in temperature between the wall of the container and the tube inside the container. Much of the experimental data and observation showed possible existence of rotating warm pockets of air floating around inside the container.

These warm pockets of air could have been created by thermals. This explanation of warm air pockets explained how the temperature, remained consistent for a few minutes, increased 3 degrees then dropped down to the temperature inside the container and then increased 3 decreased and so forth. Case14mar07a showed evidence that a "pocket" of heat moved from one thermocouple to another while spinning, creating oscillations (Figure 3.25).

Experiments showed that these "pockets of air" have moved from one thermocouple to another. There could be one or more "pockets of air." Another explanation for the movement around the container could be due convection and heat transfer.

When looking at a larger scale of air, specifically, the atmosphere that surrounds Earth, one can see temperature gradients, convention, and different fronts. Maybe a key to understanding phenomena in the container could be studying the weather. Much of weather involves chaos and the specific initial condition, much like these experiments.

It is our hope that this phenomenon will be investigated with better equipment.

**Author Contributions**
\* Carried out experiments, analyzed data (email: rsardo@nyc.rr.com), and wrote the manuscript; \*\*Designed and carried out experiments and analyzed data (email: jdbjdb@binghamton.edu).

**Acknowledgment**
We would like to thank R. Pompi for his support. Also, we would like to thank Binghamton University Department of Physics, Applied Physics, and Astronomy.

[1] T. James and D. Wales, Protonated water clusters described by an empirical valence bond potential, *J. Chem. Phy.* **122** (2005) 3061-3071

[2] N. Naguib, H. Ye, Y. Gogotsi, A. Yazicioglu, C. Megaridis, and M. Yoshimura, Observation of Water Confined in Nanometer Channels of Closed Carbon, *Nanotubes, Nano Lett.* **4** (2004) 2237-2243

[3] J. Kotz and P. Treichel, Jr., Chemistry and Chemical Reactivity, Tompson Learning, Toronto (2003) 521

[4] ibid.

[5] M. F. Chaplin, A proposal for the structuring of water, *Biophys. Chemist.* **83** (2000) 211-221

[6] J. Kotz and P. Treichel, Jr., Chemistry and Chemical Reactivity, Tompson Learning, Toronto (2003) 521

[7] F. N. Keutsch and R. J. Saykally, Water clusters: Untangling the mysteries of the liquid, one molecule at a time, *Proc. Natl. Acad. Sci. USA* **98** (2001) 10533-10540

[8] M. G. Mazza, N. Giovambattista, F. W. Starr, Dynamic Heterogeneities in Liquid Water, Phys. Rev. Lett., 96 (2006) 8031-8034

[9] M. F. Chaplin, A proposal for the structuring of water, *Biophys. Chemist.* **83** (2000) 211-221

[10] B. Chen, I. Ivanov, M. L. Klein and M. Parrinello, Hydrogen bonding in water, *Phys. Rev. Lett.* **91** (2003) 215503

[11] M. F. Chaplin, A proposal for the structuring of water, *Biophys. Chemist.* **83** (2000) 211-221

[12] H. Whiting, A new theory of cohesion applied to the thermodynamics of liquids and solids, Harvard Physics PhD Thesis, (1884).

[13] J. D. Bernal and R. H. Fowler, A theory of water and ionic solution, with particular reference to hydrogen and hydroxyl ions, *J. Chem. Phys.* **1** (1933) 515-548

[14] ibid.

[15] O. Ya. Samoilov, *Zh. Fiz. Khim.* **20** (1946) 1411

[16] ibid.

[17] M. D. Danford and H. A. Levy, The structure of liquid water at room temperature, *J. Am. Chem. Soc.* **84** (1962) 3965-3966.

[18] P. Boutron and A. Alben, Structural model for amorphous solid water, *J. Chem. Phys.* **62** (1975) 4848-4853.

[19] G. W. Robinson, C. H. Cho and J. Urquidi, Isobestic points in liquid water: Further strong evidence for the two-state mixture model, *J. Chem. Phys.* **111** (1999) 698-702.

[20] G. W. Robinson and C. H. Cho, Role of Hydration water in protein unfolding, *Biophys. J.* **77** (1999) 3311-3318

[21] R. C. Dougherty and L. N. Howard, Equilibrium structural model of liquid water: Evidence from heat capacity, spectra, density, and other properties, *J. Chem. Phys.* **109** (1998) 7379-7393

[22] ibid.

[23] ibid.

[24] M. F. Chaplin, A proposal for the structuring of water, *Biophys. Chemist.* **83** (2000) 211-221

[25] ibid.

[26] U. Kaatze and Y. Feldman, Broadband dielectric spectrometry of liquids and biosystems, *Meas. Sci. Technol.* **17** (2006) R17-R35

[27] K. X. Zhou, G. W. Lu, Q. C. Zhou, J. H. Song, S. T. Jiang and H. R. Xia, Monte Carlo simulation of liquid water in a magnetic field, *J. App. Phys.* **88** (2000) 1802-1805.

[28] J. Houghton, The Physics of Atmospheres, University Press, Cambridge (2002) 3

[29] P. F. Bernath, The spectroscopy of water vapour: Experiment, theory and applications, *Phys. Chem. Chem. Phys.* **4** (2002) 1501-1509.

[30] N. Giovambattista, S.V. Buldyrev, F. W. Starr, H. E. Stanely, Dynamic Heterogeneities in Liquid Water, AIP Conference Proceedings 708, 483 (2004)

[31] M. Conpolat, F. W. Starr, A. Scala, M. R. Sadr-Lahijany, O. Mishima, S. Haulin, H. E. Stanely, Local structural heterogeneities in liquid water under pressure, *Chem. Phys. Lett.*, **294** (1998) 9-12